\input{aipcheck}

\documentclass[
    ,final            
  ]
  {aipproc}

\layoutstyle{8x11double}

\newcommand{\lwig}{\mbox{\,\raisebox{.3ex}
    {$<$}$\!\!\!\!\!$\raisebox{-.9ex}{$\sim$}\,}}

\begin{document}

\title{String Physics at the LHC}


\author{Luis A. Anchordoqui}{
  address={
Department of Physics,
University of Wisconsin-Milwaukee,
 Milwaukee, WI 53201, USA}}

\begin{abstract}
  The LHC program will include the identification of events with
  single high-$k_T$ photons as probes of new physics. We show that
  this channel is uniquely suited to search for experimental evidence
  of TeV-scale open string theory.
\end{abstract}

\maketitle


The saga of the standard model (SM) is still exhilarating because it
leaves all questions of consequence unanswered. The most evident of
these questions concerns quantum gravity. In particular,
renormalization, the technique allowing finite predictions for
processes involving the electroweak and strong forces, fails when
gravity is taken into account. String theory is currently the only
known consistent framework to overcome the problem of
non-renormalizability and quantize
gravity~\cite{Scherk:1974ca}. Within this framework, the notion of
elementary point-particle is replaced by an extended object of
vanishing width that has two different topologies, corresponding to
open and closed strings.\footnote{The closed string sector
  automatically includes gravitation, because it contains an oscillation mode
  propagating in vacuum corresponding to a spin 2 particle which can
  be identified with the graviton.} The fundamental particles thus far
appear point-like to us because the experimental energies probed by
colliders are too small to excite the string oscillation modes. In
addition, the apparent weakness of gravity relative to the other
fundamental forces can be understood in string theory as a consequence
of the gravitational force ``leaking'' into unseen compact dimensions
transverse to a braneworld where we are
confined~\cite{Antoniadis:1998ig}.  The distance at which quantum
gravity comes into play could be then ${\cal O} (10^{-18}~{\rm
  m})$. Showld nature be so cooperative, one would expect to see a few
string states produced at the LHC.

In the perturbative regime, gauge interactions emerge as excitations
of open strings with endpoints confined on
D-branes~\cite{Polchinski:1996na}.  The basic unit of gauge invariance
for D-brane constructions is a $U(1)$ field, and so one can stack up
$N$ identical D-branes to generate a $U(N)$ theory with the associated
$U(N)$ gauge group.  Gauge bosons arise from strings terminating on
{\em one} stack of D-branes, whereas chiral matter fields are obtained
from string stretching between {\em two}
stacks~\cite{Blumenhagen:2006ci}.  Each of the two strings endpoints
carries a fundamental charge with respect to the stack of branes on
which it terminates.  Mater fields thus posses quantum numbers
associated with a bifundamental representation. None of them has three
quantum numbers of the sort we are used to: $SU(3) \times SU(2) \times
U(1)_Y$; rather they all have two quantum numbers at the expense of
introducing additional $U(1)$'s into the
theory~\cite{ant,Berenstein:2006pk}.

To develop our program in the simplest way, we will work within the
construct of a minimal model in which we consider scattering processes
which take place on the (color) $U(3)$ stack of D-branes. In the
bosonic sector, the open strings terminating on this stack contain, in
addition to the $SU(3)$ octet of gluons, an extra $U(1)$ boson
($C_\mu$, in the notation of~\cite{Berenstein:2006pk}), most simply
the manifestation of a gauged baryon number symmetry. The $U(1)_Y$
boson $Y_\mu$, which gauges the usual electroweak hypercharge
symmetry, is a linear combination of $C_\mu$, the $U(1)$ boson $B_\mu$
terminating on a separate $U(1)$ brane, and perhaps a third additional
$U(1)$ (say $W_\mu$) sharing a $U(2)$ brane which is also a
terminus for the $SU(2)_L$ electroweak gauge bosons $W_\mu^a.$ Thus,
critically for our purposes, the photon $A_\mu$, which is a linear
combination of $Y_\mu$ and $W^3_\mu$ {\em will participate with the
  gluon octet in (string) tree level scattering processes on the color
  brane, processes which in the SM occur only at one-loop level.} 

The process we consider (at the parton level) is $gg\rightarrow
g\gamma$, where $g$ is an $SU(3)$ gluon and $\gamma$ is the photon. As
explicitly calculated below, this will occur at string disk (tree)
level, and will be manifest at LHC as a non-SM contribution to
$pp\rightarrow \gamma +\ {\rm jet}$. A very important property of string
disk amplitudes is that they are completely model-independent; thus
the results presented below are robust, because {\em they hold for
  arbitrary compactifications of superstring theory from ten to four
  dimensions, including those that break supersymmetry}.

In a helicity basis, scattering amplitudes are classified according to
the number of ($\pm$) states of the external partons. The Maximum
Helicity Violating (MHV) amplitude describes the configuration with
the highest difference of ($+$) and ($-$) states, e.g. $n-2$ for a
$n$-gluon amplitude at tree level~\cite{ptmhv}. Assume that two vector bosons,
with the momenta $k_1$ and $k_2$, in the $U(N)$ gauge group states
corresponding to the generators $T^{a_1}$ and $T^{a_2}$ (here in the
fundamental representation), carry negative helicities while the other
two, with the momenta $k_3$ and $k_4$ and gauge group states $T^{a_3}$
and $T^{a_4}$, respectively, carry positive helicities.  Then the
partial amplitude for such an MHV configuration is given by~\cite{STi,STii}
\begin{eqnarray}
\label{ampl}
A(1^-,2^-,3^+,4^+) & = & 4\, g^2\, {\rm Tr}
  \, (\, T^{a_1}T^{a_2}T^{a_3}T^{a_4}) \nonumber \\ 
 & \times & {\langle 12\rangle^4\over
    \langle 12\rangle\langle 23\rangle\langle 34\rangle\langle
    41\rangle} \nonumber \\
 & \times & V(k_1,k_2,k_3,k_4)\ ,
\end{eqnarray}
where $g$ is the $U(N)$ coupling constant, $\langle ij\rangle$ are the standard
spinor products, and the Veneziano formfactor,
\begin{equation}
\label{formf}
V(k_1,k_2,k_3,k_4)=V(s,t,u)= {\Gamma(1-s)\ \Gamma(1-u)\over
    \Gamma(1+t)}\ ,
\end{equation}
is the function of Mandelstam variables, here
normalized in the string units:
\begin{equation}
\label{mandel}
s={2k_1k_2\over M_s^2},~ t={2
  k_1k_3\over M_s^2}, ~u={2 k_1k_4 \over M_s^2} \, ,
\end{equation}
with $s+t+u=0$.\footnote{We use the standard notation
  of~\cite{Mangano}, although the gauge group generators are
  normalized here in a different way, according to ${\rm
    Tr}(T^{a}T^{b})=\frac{1}{2}\delta^{ab}$.} In order to obtain the
cross section for the (unpolarized) partonic subprocess $gg\to
g\gamma$, we take the squared moduli of individual amplitudes, sum
over final polarizations and colors, and average over initial
polarizations and colors. The two most interesting energy regimes of
$gg\to g\gamma$ scattering are far below the string mass scale $M_s$
and near the threshold for the production of massive string
excitations. At low energies $(s,  t, u\ll
1)$~\cite{Anchordoqui:2007da}
\begin{equation}
\label{mhvlow}
|{\cal M}(gg\to
  g \gamma)|^2\approx  g^4Q^2C(N){\pi^4\over 4 M_s^8}(\hat s^4+\hat t^4+\hat u^4) \, ,
\end{equation}
where $C(N)=[2(N^2-4)]/[N(N^2-1)]$.
The absence of zero mass poles, at $s=0$ {\it etc.\/}, translated into
the terms of effective field theory, confirms that there are no
exchanges of massless particles contributing to this process.  On the
other hand, near the string threshold $(s \approx 1)$~\cite{Anchordoqui:2007da}
\begin{equation}
\label{mhvlow3}
|{\cal M}(gg\to g\gamma)|^2\approx
4g^4Q^2C(N){M_s^8+\hat t^4+ \hat u^4\over M_s^4(\hat s-M_s^2)^2} \, .
\end{equation}
The singularity at $\hat s=M_s^2$ needs softening to a Breit-Wigner
form, reflecting the finite decay widths of resonances propagating in
the $s$ channel. Due to averaging over initial polarizations,
Eq.(\ref{mhvlow3}) contains additively contributions from both spin
$J=0$ and spin $J=2$ gluonic Regge recurrences, created by the
incident gluons in the helicity configurations ($\pm \pm$) and ($\pm
\mp$), respectively.  The $M_s^8$ term in Eq.~(\ref{mhvlow3})
originates from $J=0$, and the $\hat t^4+ \hat u^4$ piece reflects
$J=2$ activity. Since the resonance widths are spin-dependent, 
$\Gamma^{J=0} \approx 75 \, (M_s/{\rm TeV})~{\rm GeV}$ and
$\Gamma^{J=2} \approx 45 \, (M_s/{\rm TeV})~{\rm GeV}$~\cite{widths}, 
the pole term (\ref{mhvlow3}) should be smeared as
\begin{eqnarray}
\label{mhvlow2}
|{\cal M}(gg\to g\gamma)|^2 & \simeq &
\frac{4g^4Q^2C(N)}{M_s^4} \nonumber \\
 & \times & \bigg[{M_s^8\over (\hat s-M_s^2)^2+(\Gamma^{J=0} M_s)^2}
 \nonumber \\
 &+ & {\hat t^4+\hat u^4\over (\hat s-M_s^2)^2+(\Gamma^{J=2} M_s)^2}\bigg].
\end{eqnarray}
In what follows we will take $N=3$, set $g$ equal to the
QCD coupling constant $(g^2/4\pi = 0.1)$. Before proceeding with
numerical calculation, we need to make precise the value of $Q$. If we
were considering the process $gg\rightarrow C g,$ where $C$ is the
U(1) gauge field tied to the $U(3)$ brane, then $Q = \sqrt{1/6}$ due
to the normalization condition. However, for $gg\rightarrow \gamma g$
there are two additional projections: from $C_\mu$ to the hypercharge
boson $Y_\mu$, giving a mixing factor $\kappa$; and from $Y_\mu$ onto
a photon, providing an additional factor $\cos\theta_W \ (\theta_W=$
Weinberg angle). The $C-Y$ mixing coefficient is model dependent: in
the minimal model~\cite{Berenstein:2006pk} it is quite small, around
$\kappa \simeq 0.12$ for couplings evaluated at the $Z$ mass, which is
modestly enhanced to $\kappa \simeq 0.14$ as a result of RG running of
the couplings up to 2.5~TeV.  It should be noted that in
models~\cite{ant} possessing an additional $U(1)$ which partners
$SU(2)_L$ on a $U(2)$ brane, the various assignment of the charges can
result in values of $\kappa$ which can differ considerably from
$0.12.$ In what follows, we take as a fiducial value $\kappa^2 =
0.02.$ Thus, if (\ref{mhvlow2}) is to describe $gg\rightarrow \gamma
g,$ \
\begin{equation}
Q^2= {\textstyle \frac{1}{6}} \ \kappa^2 \ \cos^2\theta_W \simeq 2.55\times
10^{-3}\ \left(\kappa^2/0.02\right)\ \ .
\label{Q2}
\end{equation}

\begin{figure}
  \includegraphics[height=.3\textheight]{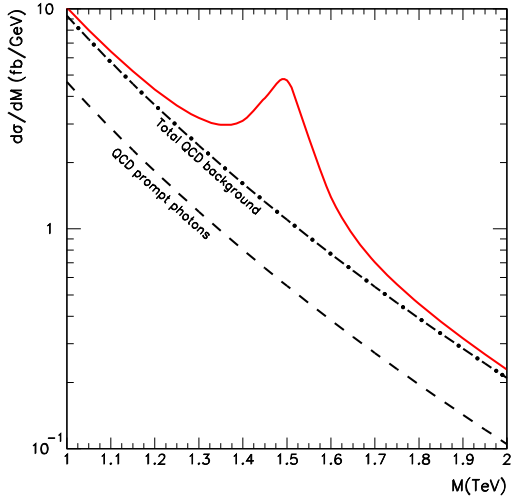}
  \caption{Invariant mass spectrum:
  $d\sigma/dM$ (units of fb/GeV) {\em vs.} $M$ (TeV) is plotted for
  the case of SM QCD background (dot-dashed line) and (first
  resonance) string signal + background (solid line). After
  minimization of misidentified $\pi^0$'s from high-$p_T$
  jets~\cite{Gupta:2007cy}, the noise is increased by a factor of
  $\sqrt{2}$ over the direct photon QCD contribution (dashed line).}
\label{fig:1}
\end{figure}
\begin{figure}
  \includegraphics[height=.3\textheight]{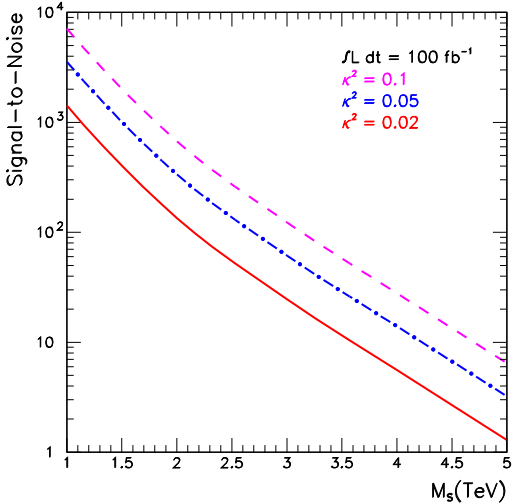}
  \caption{signal-to-noise ratio for an integrated
  luminosity of 100~fb$^{-1}$. The solid line is for $\kappa^2 =
  0.02$, the dot-dahsed line is for $\kappa^2 = 0.05$, and the dashed
  line is for an optimistic case with $\kappa^2 = 0.1$.}
\label{fig:2}
\end{figure}

One would hope that
the resonance would be visible in data binned according to the
invariant mass $M$ of the photon + jet, setting cuts on photon and jet
rapidities, $y_1,\, y_2 < y_{\rm max}$, respectively.  With the
definitions $Y\equiv {\textstyle \frac{1}{2}} (y_1 + y_2)$ and $y \equiv
{\textstyle \frac{1}{2}} (y_1-y_2)$, the cross section per interval of $M$ for
$pp\rightarrow \gamma + {\rm jet} +X$ is given
by~\cite{Anchordoqui:2008ac}
\begin{eqnarray} \frac{d\sigma}{dM} & = &  M\tau\ \sum_{ijk}\left[
\int_{-Y_{\rm max}}^{0} dY \ f_i (x_a,\, M)  \right. \ f_j (x_b, \,M ) \nonumber\\
 & \times & \int_{-(y_{\rm max} + Y)}^{y_{\rm max} + Y} dy
\left. \frac{d\sigma}{d\hat t}\right|_{ij\rightarrow \gamma k}\ \frac{1}{\cosh^2
y} \nonumber \\
& + &\int_{0}^{Y_{\rm max}} dY \ f_i (x_a, \, M) \
f_j (x_b, M) \nonumber \\ 
& \times & \int_{-(y_{\rm max} - Y)}^{y_{\rm max} - Y} dy
\left. \left. \frac{d\sigma}{d\hat t}\right|_{ij\rightarrow \gamma k}\
\frac{1}{\cosh^2 y} \right] \,,
\label{longBH}
\end{eqnarray}
where $i,j,k$ are different partons, $\tau = M^2/s$, $x_a =
\sqrt{\tau} e^{Y}$, and $x_b = \sqrt{\tau} e^{-Y}.$ The kinematics of
the scattering provides the relation $k_T = M/(2\, \cosh y)$,
which when combined with the standard cut $k_T > k_{T,
  {\rm min}}$, imposes a {\em lower} bound on $y$ to be implemented in
the limits of integration.  The $Y$
integration range in Eq.~(\ref{longBH}), $Y_{\rm max} = {\rm min} \{
\ln(1/\sqrt{\tau}),\ \ y_{\rm max}\}$, comes from requiring $x_a, \,
x_b < 1$ together with the rapidity cuts $|y_1|, \, |y_2| \le
2.4$. Finally, the Mandelstam invariants occurring in the cross
section are given by $\hat s = M^2,$ $\hat t = -{\textstyle \frac{1}{2}} M^2\ e^{-y}/
\cosh y,$ and $\hat u = -{\textstyle \frac{1}{2}} M^2\ e^{+y}/ \cosh y.$

In Figure~\ref{fig:1} we show a representative plot of this cross
section, for $M_s = 1.5$~TeV. Standard bump-hunting methods, such as
calculating cumulative cross sections, $\sigma(M_0)=\int_{M_0}^\infty
\frac{d\sigma}{dM} \, \, dM,$ and searching for regions with
significant deviations from the QCD background, may allow to find an
interval of $M$ suspected of containing a bump.  With the
establishment of such a region, one may calculate a signal-to-noise
ratio, with the signal rate estimated in the invariant mass window
$[M_s - 2 \Gamma, \, M_s + 2 \Gamma]$.  The signal-to-noise ratio is
shown in Figure~\ref{fig:2}. It is clearly seen that even for
relatively small mixing, 100~fb$^{-1}$ of LHC data could probe
deviations from SM physics associated with TeV-scale strings at a
$5\sigma$ significance, for $M_s \lwig 4~{\rm TeV}$.

\begin{theacknowledgments}
  This talk is based on several works done in collaboration with Haim
  Goldberg, Satoshi Nawata, and Tom Taylor. LAA is 
   supported by the US NSF and
  the UWM RGI.
\end{theacknowledgments}

\end{document}